\bigskip
\bigskip

\def\Z{\bf Z}
\def\[{[\![}
\def\]{]\!]}

\def\ra{\rangle}
\def\la{\langle}
\def\M{\bf M}

\baselineskip 18pt

\hfill {\bf SISSA-97/96/FM}
\smallskip
\hfill {\bf IC/96/101}
\vskip 6mm

\noindent
{\bf A Holstein-Primakoff and Dyson Realizations
for the Lie Superalgebra gl(m/n+1)}

\vskip 32pt
\noindent
{\bf Tchavdar D. Palev}\footnote*{Permanent address: 
Institute for Nuclear Research and Nuclear Energy, 1784 Sofia, 
Bulgaria; $e$-mail: tpalev@inrne.acad.bg}

\noindent
International School for Advanced Studies, via Beirut 2-4,
34123 Trieste, Italy and  \hfill\break
International Centre for Theoretical Physics, P.O. Box 586,
34100 Trieste, Italy  

\vskip 12pt

\vskip 48pt

\noindent
{\bf Abstract.} The known Holstein-Primakoff and Dyson
realizations for the Lie algebras $gl(n+1),\;n=1,2,\ldots$ in
terms of Bose operators (Okubo, S.: J. Math. Phys. {\bf 16}, 528
(1975)) are generalized to the class of the Lie superalgebras
$gl(m/n+1)$ for any $n$ and $m$. Formally the expressions are the
same as for $gl(m+n+1)$, however both Bose and $m$ Fermi
operators are involved.

\vskip 48pt

\noindent
{\bf Introduction}

\bigskip
\noindent
Recently we wrote down an analogue of the Dyson (D) and 
of the Holstein-Primakoff (H-P) realization for all
Lie superalgebras $sl(1/n)$ [1]. In the present note we extend
the results to the case of the Lie superalgebras $gl(m/n+1)$ for
any $m$ and $n$.

Initially the H-P and the D realizations were given for $sl(2)$
[2, 3]. The generalization for $gl(n)$ is due to Okubo [4].
The extension to the case of quantum algebras is available so far
only for $sl(2)$ [5] and $sl(3)$ [6]. To our best knowledge apart
from [1] other results on  H-P or D realizations for Lie
superalgebras were not published in the literature so far.

The motivation in the present work stems from the various
applications of the Holstein-Primakoff and of the Dyson
realizations in theoretical physics. Beginning with [2] and [3]
the H-P and the D realizations were constantly used in condenced
matter physics. Some other early applications can be found in the
book of Kittel [7] (more recent results are contained in [8]). For
applications in nuclear physics see [9,10] and the references
therein, but there are, ceratainly, several other publications.
In view of the importance of the Lie superalgebras for physics,
one could expect that extensions of the Dyson and of the
Holstein-Primakoff realizations to  $\Z_2$-graded
algebras may be of interest too.

We recall the H-P realization of $gl(n+1)$. The Weyl
generators $E_{AB}, \; A,B=1,\ldots,n+1$ of $gl(n+1)$ satisfy the
commutation relations:
$$
[E_{AB},E_{CD}]=\delta_{BC}E_{AD}-\delta_{AD}E_{CD}. \eqno(1)
$$
Let $b_i^\pm,\;\;n=1,\ldots,n$ be $n$ pairs of Bose creation and 
annihilation operators (CAOs),
$$
[b_i^-,b_j^+]=\delta_{ij},\quad [b_i^+,b_j^+]=[b_i^-,b_j^-]=0,
\quad i,j=1,\ldots,n.  \eqno(2)
$$
Then for any  nonnegative integer $p$, $p\in {\bf Z}_+$,
the H-P realization $\pi$ of $gl(n+1)$ is defined on the
generators as follows [4]:
$$
\eqalignno{
& \pi(E_{ij})=b_i^+b_j^-,
\quad i,j=1,\ldots,n, & (3a)\cr
& \pi(E_{i,n+1})=b_i^+\sqrt{p-\sum_{k=1}^n b_k^+b_k^-}, 
\quad 
\pi(E_{n+1,i})=\sqrt{p-\sum_{k=1}^n b_k^+b_k^-}\;b_i^-,
\quad
\pi(E_{n+1,n+1})=p-\sum_{k=1}^n b_k^+b_k^-.& (3b)\cr
}
$$
(3a) only gives
the known Jordan-Schwinger (J-S) realization of $gl(n)$ in terms
of $n$ pairs of Bose CAOs. Therefore the H-P (and also the D)
realizations are "more economical" than the J-S realization: they
allow one to express the higher rank algebra $gl(n+1)$ 
also through $n$ pairs of Bose CAOs.

Let us fix some notation. Unless otherwise stated 
$A,B,C,D=1,2,\ldots,m+n+1$ and $i,j,k,l\in
\{1,2,\ldots,m+n=M\}\equiv {\bf M}$; $[x,y]=xy-yx$,
$\{x,y\}=xy+yx$; ${\Z}_2=\{\bar{0},\bar{1}\}$; $\la A\ra=\bar{1}$,
if $A\le m$; $\la A\ra=\bar{0}$, if $A> m$.

We proceed to define $gl(m/n+1)$ in a representation independent
form.  Let $U$ be the (free complex) associative unital (= with
unity) algebra of the indeterminants
$\{E_{AB}|A,B=1,\ldots,M+1\}$ subject to the relations
$$
E_{AB}E_{CD}-(-1)^{(\la A\ra + \la B\ra)
(\la C\ra + \la D\ra)}E_{CD}E_{AB}=\delta_{BC}E_{AD}
-(-1)^{(\la A\ra + \la B\ra)(\la C\ra + \la D\ra)}E_{CB}.\eqno(4)
$$
Introduce a $\Z_2$-grading on $U$, induced from
$$
deg(E_{AB})=\la A\ra + \la B\ra. \eqno(5)
$$
Then $U$ is an (infinite-dimensional) associative superalgebra,
which is also a Lie superalgebra (LS) with respect to the 
supercommutator $\[\;,\;\]$ defined between every two homogeneous elements
$x, y \in U$ as
$$
\[x,y\]=xy-(-1)^{deg(x)deg(y)}yx. \eqno(7)
$$
Its finite-dimensional subspace
$$
{\rm lin.env.}\{E_{AB},\; \[E_{AB},E_{CD}\]  \vert
A,B,C,D=1,\ldots,m+n+1\} 
\subset U \eqno(8)
$$
gives the Lie superalgebra $gl(m/n+1)$; $U=U[gl(m/n+1)]$ is its
universal enveloping algebra.  The relations (4) are the
supercommutation relations on $gl(m/n+1)$:
$$
\[E_{AB},E_{CD}\] =\delta_{BC}E_{AD}
-(-1)^{(\la A\ra + \la B\ra)(\la C\ra + \la D\ra)}E_{CB}.\eqno(9)
$$
One can certainly define $gl(m/n+1)$ in its matrix
representation.  In that case $E_{AB}$ is a $(m+n+1)\times
(m+n+1)$ matrix with 1 on the intersection of the $A^{th}$ row
and $B^{th}$ column and zero elsewhere.

The Dyson and the Holstein-Primakoff realizations are different
embedings of $gl(m/n+1)$ into the algebra
$W(m/n)$ of all polynomials of $m$ pairs of Fermi CAOs and $n$
pairs of Bose CAOs. The precise definition of $W(m/n)$
is the following. Let $A_{i}^\pm,\;i\in \M$ be $\Z_2-$graded
indeterminats:
$$
deg(A_{i}^\pm)=\la i \ra.\eqno(10)
$$
Then $W(m/n)$ is the associative unital superalgebra
of all $A_i^\pm$, subject to the relations
$$
\[A_i^-,A_j^+\]=\delta_{ij},\quad
\[A_i^+,A_j^+\]=\[A_i^-,A_j^-\]=0. \eqno(11)
$$
With respect to the supercommutator (7) $W(m/n)$ is also a Lie
superalgebra.

From (11) one concludes that $A_1^\pm,\ldots A_m^\pm $ are Fermi
CAOs, which are odd variables; $A_{m+1}^\pm,\ldots A_{m+n}^\pm $
are Bose CAOs, which are even. The Bose operators commute with the
Fermi operators.

\smallskip
{\bf Proposition 1 (Dyson realization).} The linear map $\varphi:
gl(m/n+1) \rightarrow W(m/n)$, defined on the generators as
$$
\eqalignno{
& \varphi(E_{ij})=A_i^+A_j^-,
\quad i,j=1,\ldots,M, & (12a)\cr
& \varphi(E_{i,M+1})=A_i^+, 
\quad 
\varphi(E_{M+1,i})=(p-\sum_{k=1}^M A_k^+A_k^-)A_i^-,
\quad
\varphi(E_{M+1,M+1})=p-\sum_{k=1}^M A_k^+A_k^-.& (12b)\cr
}
$$
is an isomorphism of $gl(m/n+1)$ into $W(m/n)$ for any number $p$.

{\it Proof.} The images $\varphi(E_{AB})$ are linearly independent
in $W(m/n)$. It is straighforward to verify that they preserve the
supercommutation relations (9),
$$
\[\varphi(E_{AB}),\varphi(E_{CD})\] =\delta_{BC}\varphi(E_{AD})
-(-1)^{(\la A\ra + \la B\ra)(\la C\ra + \la D\ra)}\varphi(E_{CB}).
\eqno(13)
$$
In the intermidiate computations the following relation is useful
$$
[N,A_i^\pm]=\pm A_i^\pm,\quad {\rm where}
\quad N=\sum_{k=1}^M A_k^+A_k^-. \eqno(14)
$$

The Dyson realization defines an infinite-dimensional
representation of $gl(m/n+1)$ (for $m>0$) in the Fock space
$F(m/n)$ with orthonormed basis
$$
|K)\equiv |k_1,\ldots,k_M)={(A_1^+)^{k_1}\ldots (A_M^+)^{k_M}\over
\sqrt{k_1!\ldots k_M!}}|0\ra,\quad k_1,\ldots, k_m=0,1;
\;\; k_{m+1},\ldots, k_M\in \Z_+. \eqno(15)
$$
Let $|K)_{\pm i}$ (resp. $|K)_{i,-j}$) be a vector obtained from 
$|K)$ after a replacement of $k_i$ with $k_i\pm 1$ 
(resp. $k_i \rightarrow k_i+1,\; k_j \rightarrow k_j-1$). The
transformations of the basis (15) under tha action of the CAOs
read:
$$
A_i^+|K)=(-1)^{\la i \ra (k_1+\ldots +k_{i-1})}
\sqrt{1+(-1)^{\la i \ra}k_i}\;|K)_i,\quad 
A_i^-|K)=(-1)^{\la i \ra (k_1+\ldots +k_{i-1})}
\sqrt{k_i}\;|K)_{-i}. \eqno(16)
$$
As a consequence one obtains the transformations of the $gl(m/n+1)$
module $F(m/n)$:
$$
\eqalignno{
& \varphi(E_{i,M+1})|K)=(-1)^{\la i \ra (k_1+\ldots +k_{i-1})}
  \sqrt{1+(-1)^{\la i \ra}k_i}\;|K)_i,   & (17a) \cr
& \varphi(E_{M+1,i})|K)=(-1)^{\la i \ra (k_1+\ldots +k_{i-1})}
(p+1-\sum_{j=1}^M k_j)\sqrt{k_i}|K)_{-i},   & (17b) \cr
& \varphi(E_{M+1,M+1})|K)= (p+1-\sum_{j=1}^M k_j)|K),&(17c)\cr
& \varphi(E_{ii})|K)=k_i|K),   & (17d) \cr
& \varphi(E_{ij})|K)=(-1)^{\la i \ra (k_1+\ldots +k_{i-1})
  +\la j \ra (k_1+\ldots +k_{j-1}) }
  \sqrt{k_j(1+(-1)^{\la i \ra}k_i)}\;|K)_{-j,i},
  \quad i<j, & (17e) \cr
& \varphi(E_{ij})|K)=(-1)^{\la i \ra (k_1+\ldots +k_{i-1}+1)
  +\la j \ra (k_1+\ldots +k_{j-1}) }
  \sqrt{k_j(1+(-1)^{\la i \ra}k_i)}\;|K)_{-j,i},
  \quad i>j, & (17f) \cr  
}
$$
If $p$ is not a positive integer, $p\notin {\bf N}$, $F(m/n)$
is a simple $gl(m/n+1)$ module. For any positive integer
$p$, $p\in {\bf N}$, the representation
of $gl(m/n+1)$ in $F(m/n)$ is indecomopsible. The subspace
$$
F(p;m/n)_{inv}=lin.env.\{|K)\;|k_1+\ldots+k_M>p\}\subset
F(m/n) \eqno(18)
$$
is an  infinite-dimensional subspace, invariant with respect to
$gl(m/n+1)$. The factor spaces
$$
F(p;m/n)_0\equiv F(m/n)/F(p;m/n)_{inv}=
lin.env.\{|K)\;|p\le k_1+\ldots+k_M \},\quad p=1,2,\ldots,
\eqno(19)
$$
are finite-dimensional irreducible $gl(m/n+1)-$modules.

The advantage of the Dyson realization (12) is its simplicity.
Its disadvantage - the Fock representation of $gl(m/n+1)$ is not
unitarizable. The latter, the representation to be unitarizable,
is usually required for physical reasons.  We recall that a
representation $\varphi$ of a (super)algebra $L$ in a Hilbert space
$V$ is unitarizable with respect to an antilinear antiinvolution
$\omega: L\rightarrow L$ and a scalar product $(\;,\;)$ in $V$,
if
$$
(\varphi(a)x,y)= (x,\varphi(\omega(a))y),\quad \forall a\in L,\;\;
\forall x,y\in V.
\eqno(20)
$$
The Dyson representation in $F(m/n)$ is not unitarizable with
respect to the "compact" antiinvolution
$$
\omega(E_{AB})=E_{BA},\quad A,B=1,\ldots,M+1. \eqno(21)
$$
The factor-modules $F_0(p;m/n),\;\;p\in {\bf N}$, however do carry
unitarizable representations for any $p\in {\bf N}$. In order to
show this it is convenient to introduce a new basis within each
$F_0(p;m/n)$, which we postulate to be orthonormed:
$$
|K\ra=\sqrt{(p-\sum_{j=1}^M k_j)!}\;|K). \eqno(22)
$$
In this basis the transformation relations  (17) read:
$$
\eqalignno{
& \varphi(E_{i,M+1})|K\ra=(-1)^{\la i \ra (k_1+\ldots +k_{i-1})}
  \sqrt{(1+(-1)^{\la i \ra}k_i)(p-\sum_{j=1}^M k_j)}\;|K\ra_i,&(23a)\cr
& \varphi(E_{M+1,i})|K\ra=(-1)^{\la i \ra (k_1+\ldots +k_{i-1})}
\sqrt{k_i(p+1-\sum_{j=1}^M k_j)}|K\ra_{-i},   & (23b) \cr
& \varphi(E_{M+1,M+1})|K\ra= (p+1-\sum_{j=1}^M k_j)|K\ra,&(23c)\cr
& \varphi(E_{ii})|K\ra=k_i|K\ra,   & (23d) \cr
& \varphi(E_{ij})|K\ra=(-1)^{\la i \ra (k_1+\ldots +k_{i-1})
  +\la j \ra (k_1+\ldots +k_{j-1}) }
  \sqrt{k_j(1+(-1)^{\la i \ra}k_i)}\;|K\ra_{-j,i},
  \quad i<j, & (23e) \cr
& \varphi(E_{ij})|K\ra=(-1)^{\la i \ra (k_1+\ldots +k_{i-1}+1)
  +\la j \ra (k_1+\ldots +k_{j-1}) }
  \sqrt{k_j(1+(-1)^{\la i \ra}k_i)}\;|K\ra_{-j,i},
  \quad i>j, & (23f) \cr  
}
$$
It is straightforward to check that (20) hold with respect
to the antiinvolution (21). Hence the reresentation of $gl(m/n+1)$
is unitarizable withing every space $F_0(p;m/n),\;\;p\in {\bf
N}$. The next proposition is closely related to the result we
have just obtained.

\smallskip
{\bf Proposition 2 (Holstein-Primakoff realization).} 
The linear map $\pi: gl(m/n+1) \rightarrow W(m/n)$, defined
on the generators as
$$
\eqalignno{
& \pi(E_{ij})=A_i^+A_j^-,
\quad i,j=1,\ldots,M, & (24a)\cr
& \pi(E_{i,M+1})=A_i^+\sqrt{p-\sum_{j=1}^M A_j^+A_j^-}, 
\; 
\pi(E_{M+1,i})=\sqrt{p-\sum_{k=1}^M A_k^+A_k^-}\;A_i^-,
\;
\pi(E_{M+1,M+1})=p-\sum_{k=1}^M A_k^+A_k^-.& \cr
& &(24b)\cr
}
$$
is an isomorphism of $gl(m/n+1)$ into $W(m/n)$ for any 
positive integer $p$.

{\it Proof.} Acting with the $gl(m/n+1)$ generators on the basis (15)
one obtaines the same transformation relations (23) with the only
difference that everywhere in (23) $|K\ra$ have to be replaced
with $|K)$. The proof can be carried out also purely
algebraicaly, using the supercommutation relations (11). To this
end the following formula is useful:
$$
f(N)A_i^\pm=A_i^\pm f(N\pm 1),\quad N=\sum_{j=1}^M A_j^+A_j^-,
\eqno(25)
$$
where $f(z)$ is any (analitical) function in $z$.

The representation $\pi$ is defined in the entire Fock space.
Observe that with respect to $\pi(E_{AB}),\;\;A,B=1,\ldots,M+1$,
the Fock space resolves into a direct sum of two invariant (and
moreover irreducible) subspaces (which was not the case with the
Dyson representation):
$$
 F(p;m/n)_0=lin.env.\{|K)\;|p\le k_1+\ldots+k_M\},\;
 F(p;m/n)_{inv}=lin.env.\{|K)\;|k_1+\ldots+k_M>p\}.\eqno(26)
$$
This propertiy is due to the factors $\sqrt{p-\sum_{j=1}^M k_j}$ and 
$\sqrt{p+1-\sum_{j=1}^M k_j}$
in (23a) and (23b), respectively.

In the case $m=0$ the Holstein-Primakoff realization (24) reduces
to the Holstein-Primakoff realization (3) of $gl(n+1)$ in terms of
only Bose operators. Taking in (24) all $A_i^\pm$ to be Bose
CAOs, one obtains the H-P realization of $gl(m+n+1)$.  The case
$n=0$ yields a Fermi realization of the Lie superalgebrs
$gl(m/1)$. Its restriction to $sl(m/1)$ coincides with the results
announced in [1].

Let us note in conclusion that explicit expressions for 
all finite-dimensional irreducible representations of $gl(m/1)$ 
and a large class of representations of $gl(m/n+1)$ are
available [12]. They have been generalized also to the quantum
case [13]. The formulae are however extremely involved. The Dyson
and the Holstein-Primakoff representations lead to a small part of
all representations. Their description is however simple and it
is given in familiar for physics Fock spaces.

\vskip 24pt
\noindent
{\it Acknowledgments.}
The author is thankful to Prof. Randjbar-Daemi for the kind
hospitality at the High Energy Section of ICTP. It is a pleasure
to thank Prof. C. Reina for making it possible to visit the
Section on Mathematical physics in Sissa,  where most of the
results were obtained. The work was supported by the Grant
$\Phi-416$ of the Bulgarian Foundation for Scientific Research.

\vskip 24pt
\noindent
{\bf References}

\vskip 12pt
\settabs \+  [11] & I. Patera, T. D. Palev, Theoretical 
   interpretation of the experiments on the elastic \cr 
   
\+ [1] & Palev, T.D.: An analogue of Holstein-Primakoff and Dyson
         realizations for Lie superalgebras. \cr
\+     & The Lie superalgebra $sl(1/n)$. Preprint IC/96/91 \cr

\+ [2] & Holstein, T., Primakoff, H.: Field dependence of the intrinsic
         domain magnetization of a ferromagnet. \cr
\+     & Phys. Rev. {\bf 58}, 1098-1113 (1940) \cr            

\+ [3] & Dyson, F. J.: General theory of spin-wave interactions.   
           Phys. Rev. {\bf 102}, 1217-1230 (1956)\cr

\+ [4] & Okubo, S.: Algebraic identities among $U(n)$ infinitesimal
         generators. J. Math. Phys. {\bf 16}, 528-535 (1975)\cr

\+ [5] & Chaichian, M., Ellinas, D., Kulish, P. P.: Quantum algebra
         as the dynamical symmetry of the deformed \cr
\+	   & Jaynes-Cumminis model. Phys. Rev. Lett. 
         {\bf 65}, 980-983 (1990) \cr

\+ [6] & da-Providencia, J.: Mean-field generation of the classical
         $q$-deformation of $su(3)$. \cr
\+	   &    J. Phys. A {\bf 26}, 5845-5849 (1993)\cr

\+ [7] & Kittel, C.: Quantum Theory of Solids, Willey, 
          New York (1963) \cr

\+ [8] & Caspers, W. J.: Spin Systems, World Sci. Pub. Co., Inc.,
          Teanek, NJ (1989) \cr

\+ [9] & Klein, A., and Marshalek, E. R.: Boson realizations of Lie
         algebras with applications to nuclear physics. \cr
\+     & Rev. Mod. Phys. {\bf 63}, 375-558 (1991) \cr

\+ [10] & Ring, P., Schuck, P.: The Nuclear Mani-Body Problem, 
         Springer-Verlag, New York, Heidelberg, Berlin.\cr

\+ [11] & Kac, V. G., Representations of classical Lie
          superalgebras.  \cr
\+      &  Lect. Notes Math. {\bf 676}, 597-626 (1978) \cr

\+ [12] &  Palev, T. D.: Irreducible finite-dimensional
           representations of the Lie superalgebras $gl(n/1)$  \cr
\+      &  in a Gelfand-Zetlin basis.  Funkt. Anal. Prilozh.
           {\bf21}, \#3, 85-86 (1987) (In Russian);  \cr
\+	    &  Funct. Anal. Appl.
	       {\bf 21}, 245 (1987) (English translation)\cr

\+      &  Palev, T. D.: Essentially typical
           representations of the Lie superalgebras $gl(n/m)$  \cr
\+      &  in a Gelfand-Zetlin basis. Funkt. Anal. Prilozh.
           {\bf23}, \#2, 69-70 (1989) (In Russian);  \cr
\+	    &  Funct. Anal. Appl.
	       {\bf 23}, 141-142 (1989) (English translation)\cr

\+ [13] & Palev, T. D., Tolstoy, V. N.: Finite-dimensional 
          irreducible representations of the quantum \cr
\+		& superalgebra $U_q[gl(n/1)]$. Comm. Math. Phys.
          {\bf 141}, 549-558 (1991)\cr

\+      & Palev, T. D., Stoilova, N. I., Van der Jeugt, J.: 
          Finite-dimensional representations of the quantum \cr
\+		& superalgebra $U_q[gl(n/m)]$ and related $q$-identities.   
          Comm. Math. Phys. {\bf 166}, 367-378 (1994)\cr

\end